\documentclass{llncs}
    \usepackage{graphicx}
    \usepackage{mmap}
    \usepackage{wrapfig}
\usepackage{amssymb}       
\usepackage[latin1]{inputenc}
\usepackage[english]{babel}
\usepackage[T1]{fontenc}

    \usepackage[bookmarks=false, linkcolor=black, citecolor=black,
      urlcolor=black,colorlinks=true, breaklinks=true, bookmarksopen=false,
      pdftitle={Understanding the Learners' Actions when using Mathematics Learning Tools},
      pdfauthor={Paul Libbrecht, Sandra Rebholz, Daniel Herding, Wolfgang Müller, Felix Tscheulin},
      pdfkeywords={teaching mathematics, learning tools, traces, learning log, problem solving solution paths, privacy, learning management systems},
     ]{hyperref}

\begin{document}
\title{Understanding the Learners' Actions \\when using Mathematics Learning Tools}
\titlerunning{Understanding the Learners' Actions...}  
%
  \author{Paul Libbrecht${}^1$, Sandra Rebholz${}^2$, Daniel Herding${}^3$, \\ Wolfgang Müller${}^2$, Felix Tscheulin${}^2$
     \institute{${}^1$ Institute for Mathematics and Informatics, Karlsruhe University of Education,\\ 
     ${}^2$Media Education and Visualization Group, Weingarten University of Education,\\
     ${}^3$Computer-Supported Learning Research Group, RWTH Aachen University\\
     Germany
     }}
\maketitle              

\begin{abstract}
The use of computer-based mathematics tools is widespread in learning. Depending on the way that these tools assess the learner's solution paths, one can distinguish between automatic assessment tools and semi-automatic assessment tools. Automatic assessment tools directly provide all feedback necessary to the learners, while semi-automatic assessment tools involve the teachers as part the assessment process. They are provided with as much information as possible on the learners' interactions with the tool. 

How can the teachers know how the learning tools were used and which intermediate steps led to a solution?
How can the teachers respond to a learner's question that arises while using a computer tool?
Little is available to answer this beyond interacting directly with the computer  and performing a few manipulations to understand the tools' state.

This paper presents SMALA, a web-based logging architecture that addresses these problems by recording, analyzing and representing user actions. While respecting the learner's privacy, the SMALA architecture supports the teachers by offering fine-grained representations of the learners' activities as well as overviews of the progress of a classroom.

\end{abstract}
\section{Learners' Actions and the Perception of Teachers}

\begin{wrapfigure}{r}{50mm}\vspace{-10mm}\includegraphics[width=50mm]{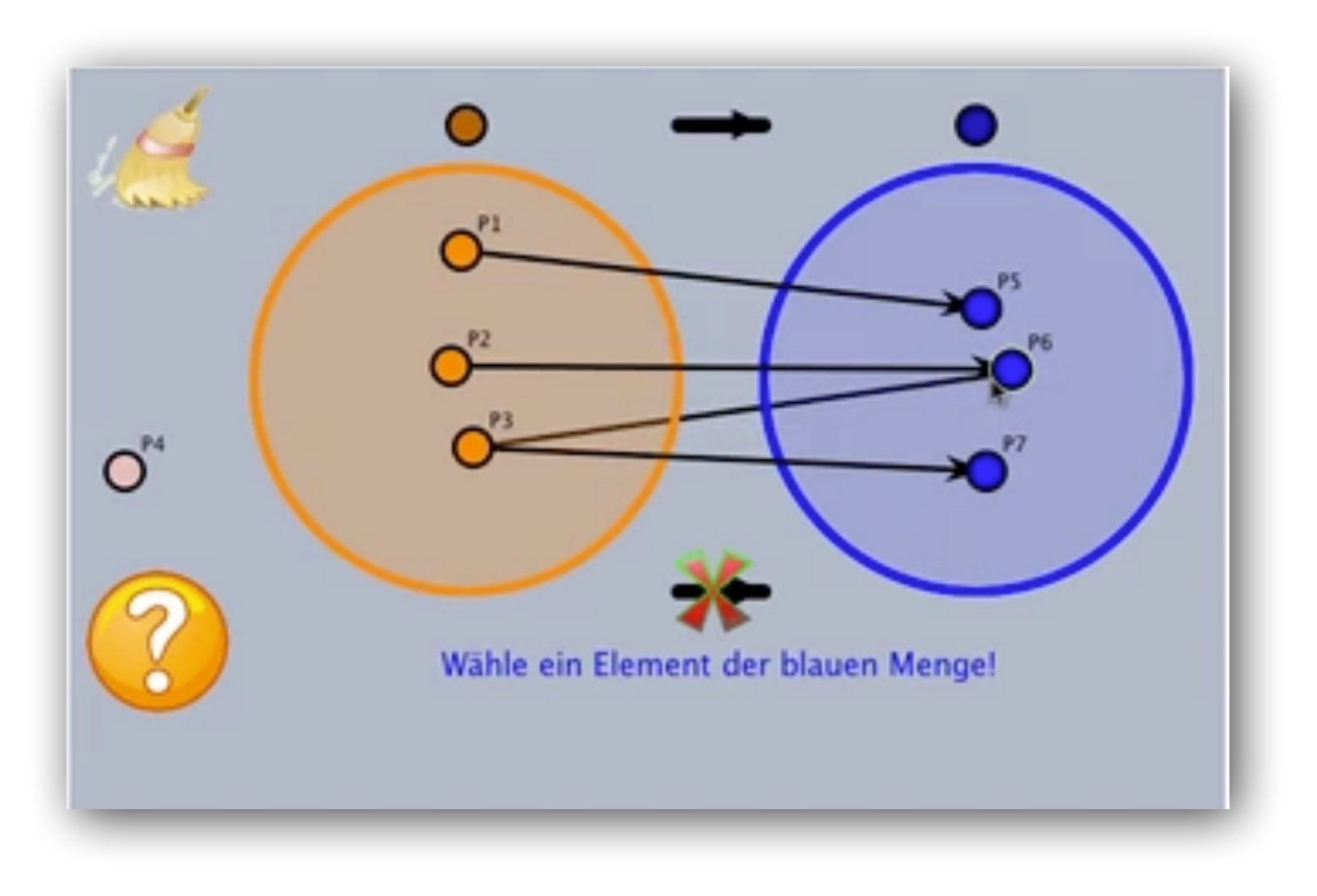}\vspace{-10mm}\end{wrapfigure}
Interactive learning tools offer rich possibilities to students when learning mathematics. On the one hand, they offer unprecedented possibilities to explore dynamic representations of the mathematical concepts: they allow students to discover the domain's objects and the rules that hold between them as much or as little as they want. Examples of such learning tools include function plotters and dynamic geometry systems. Another example is pictured on the right: the tool Squiggle-M~\cite{Fest-Feedback} is used to explore relations between finite sets in order to learn about functions, injectivity, and surjectivity.

\begin{wrapfigure}{l}{80mm}\vspace{-2mm}\includegraphics[width=80mm]{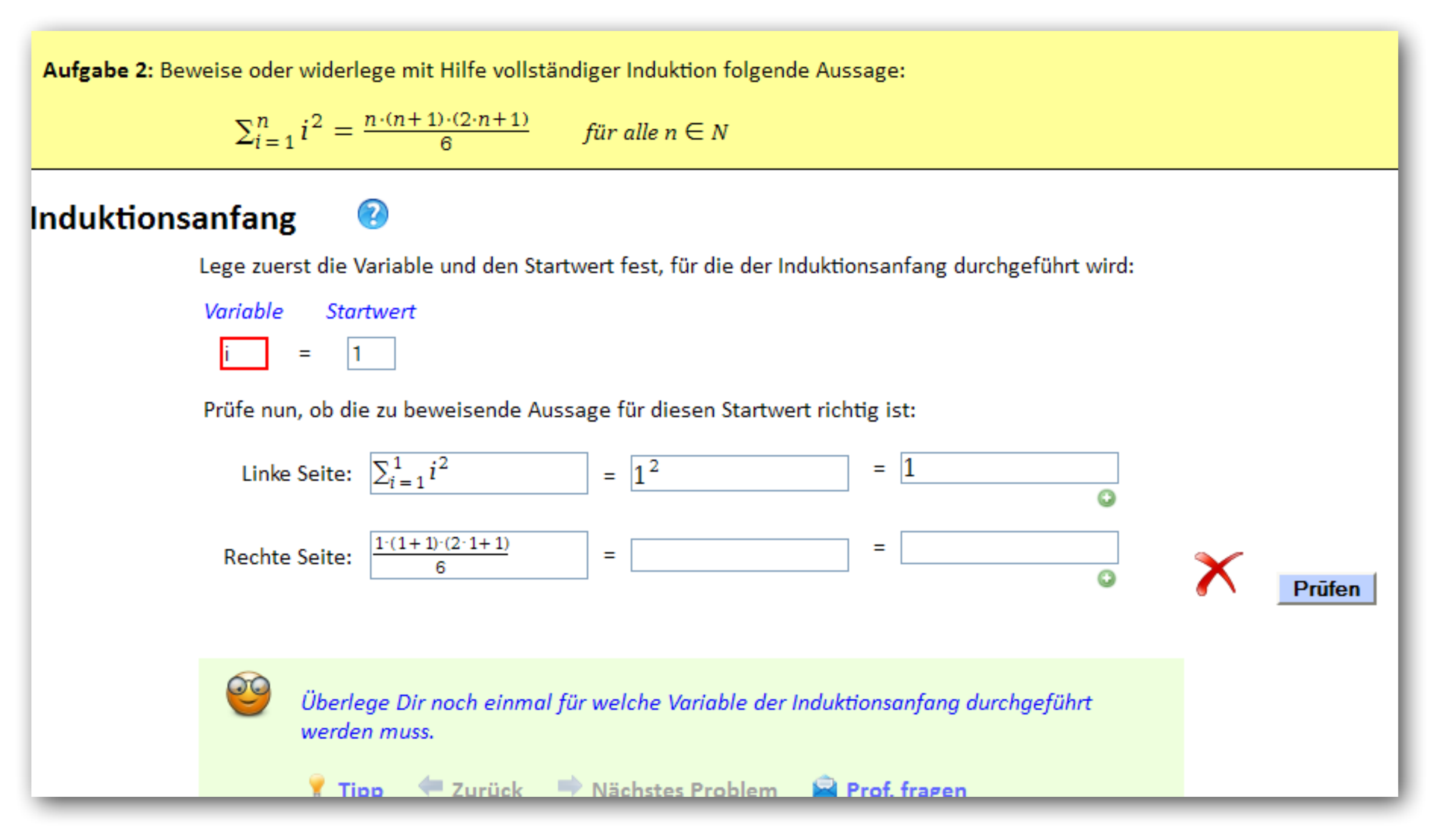}\vspace{-8mm}\end{wrapfigure}On the other hand, interactive tools for learning mathematics support the automated corrections of a broad range of typical exercises which allow learners to train and obtain the routine that stabilizes their mastery of the concepts. Trainers of this sort include the domain-reasoner powered exercises of ActiveMath~\cite{gogmkm09} and  many of the cognitive tutors~\cite{Cognitive-TUtors-Big-City-IJAIED-1997}. Another example is pictured on the left, that of ComIn-M~\cite{Rebholz-Zimmermann-ComInM-Elearning-Baltics-2011}: it allows the learners to apply the classical workflow of proof by induction to proofs about number-theoretical formul{\ae}.

Whenever computer tools are used in learning scenarios, the teacher plays a central role: he\footnote{In this paper, we shall use the feminine for the learner and the masculine for the teachers even though we mean both genders for both roles.} explains the concepts by using representations and operations that the learner can also find in the learning tools. He invites the learners to use the learning tool: For example, he can indicate that a few exercises of the assignment sheet are to be done using the learning tools which they can find in the learning management systems.
These usage incentives are not able to ensure the quality of the usage that enhances what the learners have acquired. Indeed, all sorts of risks appear when learning tools are used:

\vspace{-2mm}\begin{itemize}
\item An overwhelming cognitive load when introducing the learning tools (caused by the amount of technical details, for example).
\item Too steep a learning curve to use the tool (when the students need to concretely apply the mathematical knowledge).
\item Choice of exercises that lead to frustration due to unachieved exercises (for mathematical, technical, or other reasons).
\end{itemize}\vspace{-2mm}

These risks are challenges to teachers -- the issue is underlined as insufficiently addressed in the report~\cite{Artigue-Challenges-Basic-Math-Education-Unesco}: {\it The recurrent difficulties} [...] {\it call into question both the design of resources and the processes of their dissemination}.

To assess these risks, teachers must understand the students' usage of the learning tools. 
But providing detailed logs of the students' actions is not sufficient since the details may be overwhelming. The vast amount of data that could be produced by logging usually prevents the teachers using such a source to deduce any useful information about the students' learning processes.

\paragraph{User Requirements}
Approaches and tools are required that allow the efficient analysis and understanding of such log data. Learning analytics is the domain investigating me\-thods and tools to support this.\footnote{Learning analytics is defined by G. Siemens as ``the use of intelligent data, learner-produced data, and analysis models to discover information and social connections, and to predict and advise on learning.'' in \url{http://www.elearnspace.org/blog/2010/08/25/what-are-learning- analytics}.}

Teachers should be able to capture the overall progression of the learners in a class; being informed on the successes and failures for each exercise and each skilled acquisition aimed at. This has the potential to guide the lecture's content for such adjustments as revising a conceptual error commonly found or demonstrating a manipulation in more detail with the same concepts and representations of the learning tools.

In addition, we consider it important that the teachers complement the automatic assessment capabilities of the learning tools, making it a semi-automatic assessment~\cite{BHKMZ11}.
Teachers should be able to see the detailed inputs a learner has made and the precise automatic feedback she received when requested to help. 
In the case of requesting help while using the learning tools, the learners should be able to formulate a request for help linked to the list of events that occurred until they needed assistance so that the teacher can analyze what the learner did and suggest effectively what actions to take next (in both computer-technical and mathematical terms).

In order to realize these objectives, teachers need a logging infrastructure which is the focus of this contribution.

\paragraph{Technical Requirements}
The purpose of this research aims at serving teachers in universities to support the use of richly interactive learning tools, typically of client-based applet-like tools which have not been designed with action logging in mind.
We aim to insert an architecture for logging into the widespread infrastructure of learning management systems (LMS): those university-central systems that are used for coordination of courses and which each student regularly visits. 
We aim our development to not require an LMS change; indeed we have often met such a desire to be impossible to satisfy in university wide learning management systems.

Thus, one of the basic technical challenges is that of enabling teachers, who are privileged users of the LMS, to provide their learners with methods to start the learning tools from the LMS so that identified logs are received.

Other technical challenges revolve around the display of relevant information about the learning actions to the teachers in a way that is easily accessible and navigable.
The learning tools, the LMSs, and the log views should be web-based and allow the servers to recognize the identity of learners and teachers.

\paragraph{At the Edge of Privacy}

A major concern of learning analytics is the set of regulations about the users' privacy. Indeed, the usage of such a monitoring tool may be turned into a powerful watching tool if not used carefully. Moreover, we acknowledge that a part of the students we have met are bothered using a tool where each of the attempted solution paths are always visible.

In comparison to log-views that irreversably show all steps of the problem solving process, 
the classroom based usage where teachers and assistants can come and see the current state of the learning tool generally allows the student to cancel (and thus hide) erroneous steps that are irrelevant to a teacher question. 

Thus, to support some free choice of disclosure of the students, we set forth the following principles that respect Germany's and  EU's laws on privacy:


\vspace{-3mm}\begin{itemize}
\item The log of a session is only associated to an identifiable person when that individual expressly consents to being identified, i.e. when requesting help.
\item The students always have the possibility to opt-out of the log collection.
\item The information recorded is transparent to the learner.
\end{itemize}\vspace{-3mm}

These principles do not prevent all sessions of the same learner being grouped together, as long as it remains impossible for a teacher to associate a person with the log view of a session. As we shall describe below, this will be addressed by presenting the learners' {\it pseudonym}, a barely readable number derived from the name. We acknowledge that teachers would still be able to track regularly by remembering the pseudonym (or by many other means), but we explicitly warn the teachers that such is the start of illicit monitoring.

These principles respect the privacy laws in a same manner as the widespread Twitter or Facebook widgets in web-pages: they do not require a supplementary privacy agreement by the students since the log information that is collected and made available does not contain personal information.

Again similarly to these services, the privacy disclosure is agreed upon when the learners explicitly decide to do so: in the case of such services, this implies registrations, in the case of SMALA, it is when requesting help.
\section{State of the Art in Logging User Actions}
Logging users' actions can be done in multiple ways; in this section we outline existing approaches that are described in the current research literature. They revolve around two axes: the methods to integrate learning tools in learning management systems and the log-collection and log-view approaches.

\vspace{-3mm}\subsection{Standards for Integrating Learning Tools}

The widespread SCORM packaging standard\footnote{The Shareable Content Object Reference Model standard emerged from ADLnet about 10 y. ago. See \url{http://www.adlnet.gov/Technologies/scorm/default.aspx}.} allows makers of learning tools to bundle a sequence of web pages that use an API for communication with the LMS.  
A more recent derivative of SCORM is Common Cartridge, which extends it with IMS Learning Tools Interoperability IMS-LTI\footnote{The Learning Tools Interoperability specification is an emerging standard, see  \url{http://www.imsglobal.org/toolsinteroperability2.cfm}.}: this specification allows the teacher of a module in an LMS to publish enriched forms of links which carry the authentication.

Both SCORM and Common Cartridge provide basic infrastructures for the web integration of learning tools. However, their logging capabilities in terms of collecting and analysing usage data is very limited. Supported log views typically show tables or counts inside the LMS. For a teacher to be able to evaluate the progress of one learner, he needs to see a less abstract view, more resembling the view the learner had when performing a learning activity.

Multiple other standards have been realized to provide single-sign-on infrastructures: \cite{ReverseOAuth} describes many of these infrastructures and conclude with the proposal of yet another approach to authentication.

\vspace{-3mm}\subsection{Research Around Log Collections}\vspace{-1mm}

There are various approaches for collecting logs of user activities and making them available via suitable views. In the following, we will characterize the logging approaches by the level of detail of the collected data, the semantic content that is available, and the analysis capabilities that are offered.
\paragraph{Jacareto:} The most detailed approach for log collection is to record each of the user's actions and present these as a {\it replay} of the learning tool's user interface. This approach has been investigated in a software project called Jacareto \cite{Spannagel-Glaser-Zikuda-Schroeder}: apart from input events such as mouse clicks in the learning application tool-specific {\it semantic events} are stored in a recording file. Once the teacher obtains the record file, he can analyze the solution process qualitatively -- either by looking at a hierarchical view of the events, or by replaying the events and observing the learning application. However, quantitative analysis of a large number of recordings is not supported, and remote logging is not yet available. 

\paragraph{FORMID:} The research project FORMID \cite{Gueraud-Cagnat-ECTEL06} aims at applying learning scenarios and view logs based on these scenarios. Each scenario is implemented as a script that specifies a sequence of learning activities to be followed by learners. Additional monitoring facilities enable the teacher to observe learning activities and provide support in real time. In this approach, the semantic content of the logging data is quite rich, and sufficient analysis capabilities are offered for assessing the learners' progression but only within the given scenario. 

\paragraph{LOCO-Analyst:} The approach of LOCO-Analyst \cite{LOCOanalys-IJCEELL} intends to support the teacher in analyzing learning processes in order to optimize and revise content elements in online learning courses. For this reason, LOCO-Analyst focuses on online learning activities such as text reading and obtaining scores when solving multiple-choice exercises. Based on semantic web technologies, LOCO-Analyst enhances logging data by semantic annotation and provides various analysis services and graphical representations of the collected data. As opposed to Jacareto, the level of detail of the logging data is quite low: it mostly considers high-level events such as page views or successful exercise solutions.


\paragraph{Log Repositories:} Learners' activities logs are the basic input for the PSLC DataShop\footnote{For the PSLC DataShop see \url{https://pslcdatashop.web.cmu.edu/}.} and for a similar initiative in Kaleidoscope~\cite{MelisMcLarenSolomonECTEL2008}. These initiatives are infrastructures for collecting logs of learning for their massive evaluations for such purposes as data-mining to experimentally measure e-learning theories. These logging repository initiatives are researcher-oriented: they  bring together quantities of logs in order to formulate hypotheses about the learners' actions. They are not applicable for teachers, and often do not work on live streams of data.

\subsection{Conclusions from Literature Review}\vspace{-1mm}


From this review of the existing literature we conclude that no current standard nor widespread learning management system offers sufficient support for the deployment of logging-enabled learning tools run on the client or even for any interaction-rich learning tools. 

We also conclude that the approaches proposed by state-of-the-art research initiatives provide interesting logging features but only few approaches are flexible enough to log semantically-rich events in an adequate level of detail. Customizable views that offer an efficient display of the logging data enhanced by flexible analysis capabilities are still subject to ongoing research.

\begin{figure}\vspace{-5mm}\begin{center}
\includegraphics[width=11cm]{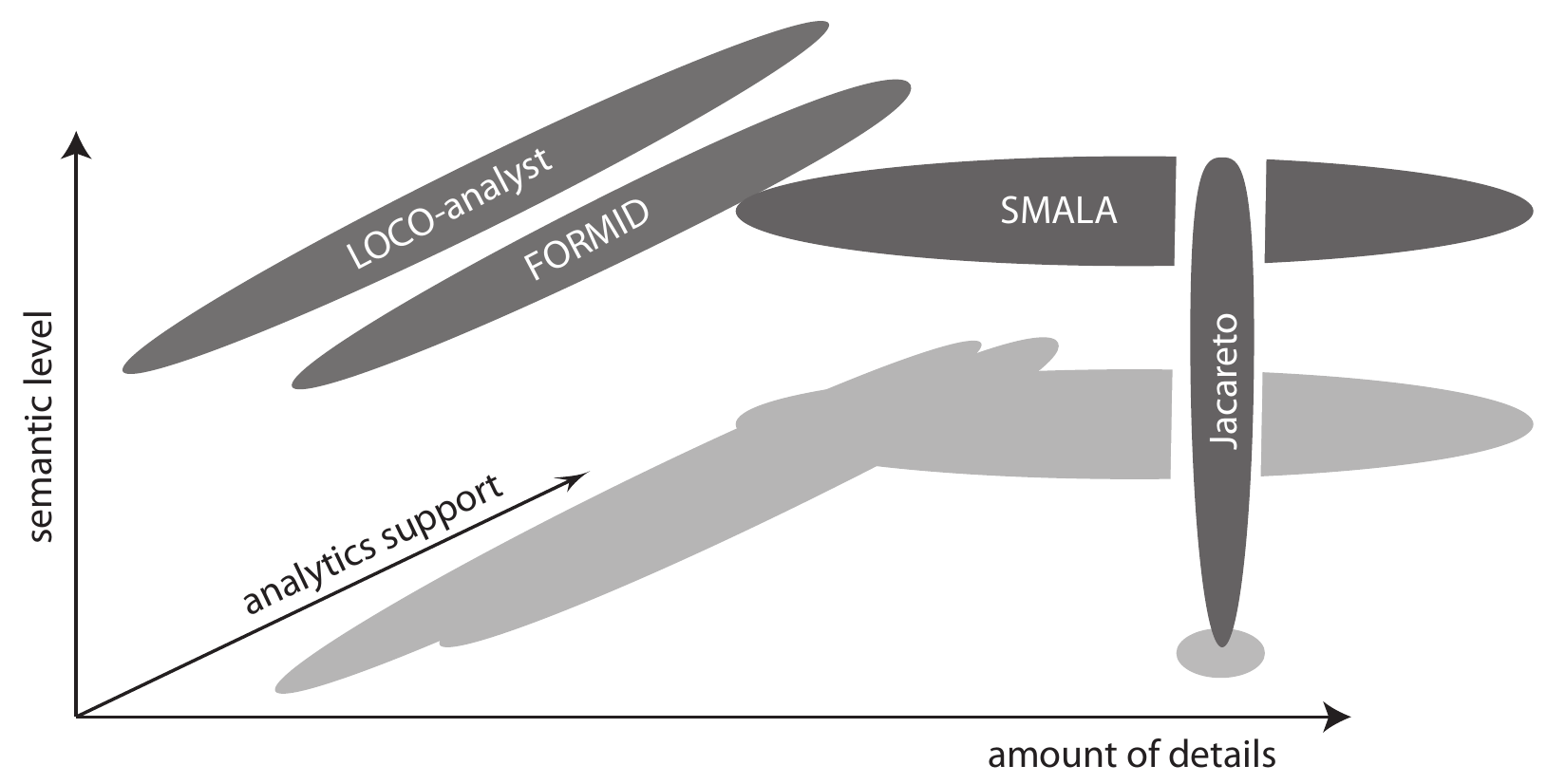}
\end{center}
\vspace{-0.5cm}\caption{The various logging systems along three dimensions}\label{fig:log-dimensions}\vspace{-0.5cm}
\end{figure}

Figure~\ref{fig:log-dimensions} summarizes how the existing logging approaches can be assigned to the dimensions {\it amount of detail}, {\it semantic level}, and {\it analytics support}. As shown in the diagram, the SMALA architecture that we present in the following chapter aims at high values in all three dimensions, with an intent to cover different amounts of details, and thus represents a new category of logging system.

\section{The SMALA Architecture}
\begin{figure}\begin{center}
\includegraphics[width=11cm]{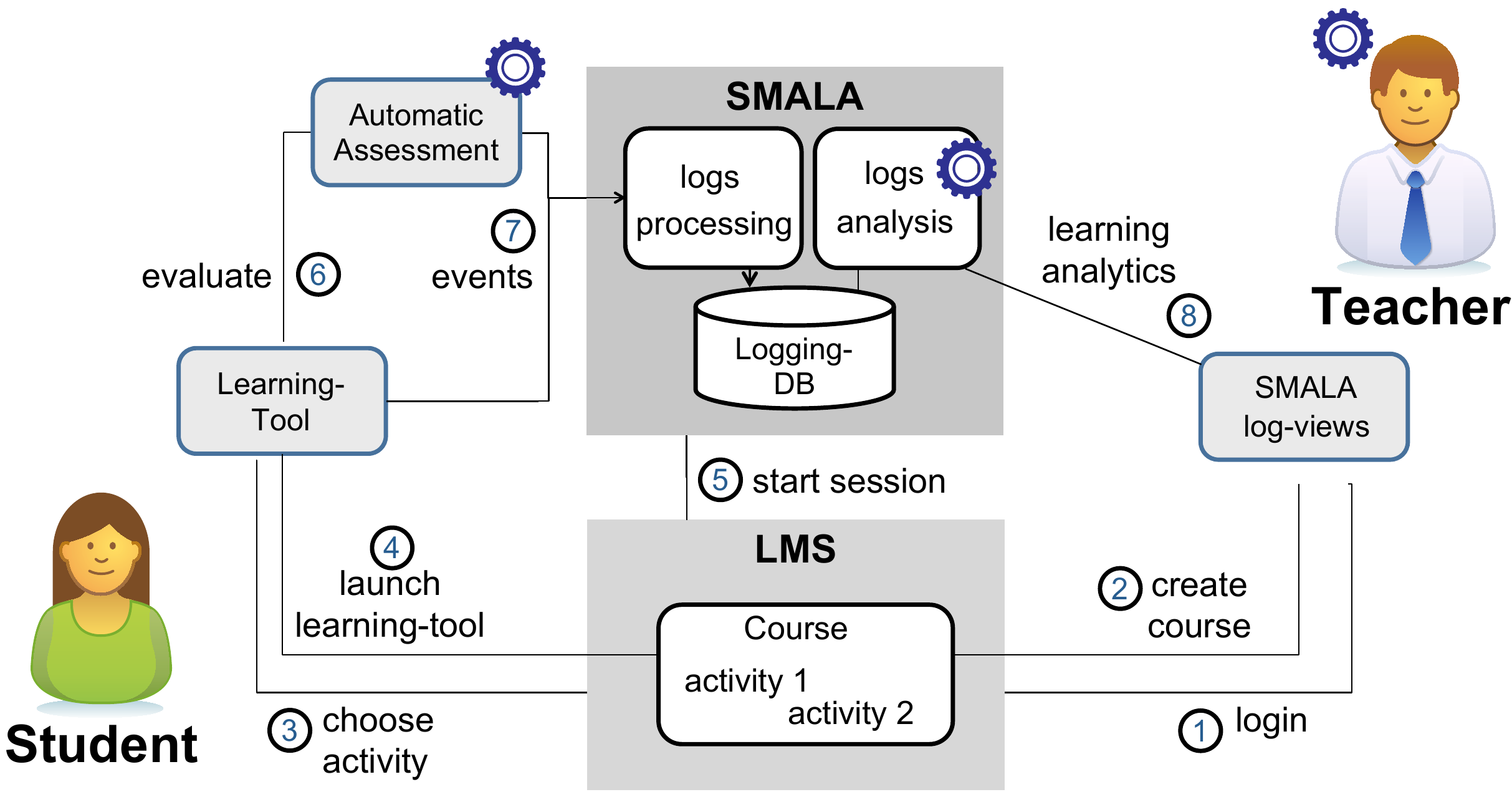}
\end{center}
\vspace{-0.5cm}\caption{The SMALA architecture. Gear wheels indicate actors and components with analytical reasoning.}
\label{fig:smala-architecture}
\end{figure}

The contribution of our paper is an architecture for logging and analyzing learners' activities and solution paths. SMALA, {\sc Sail-M}'s Architecture for Learning Analytics, responds to the requirements stated above and is realized as a service-oriented web application. Figure~\ref{fig:smala-architecture} depicts the SMALA system architecture where the sequence of components in the creation, deployment, initialization, usage, assessment, logging, and log observation is numbered.

SMALA's core component is the SMALA logging service that is responsible for receiving events from the learning tools and storing them persistently in the logging database. Authorized teachers can retrieve the recorded events from the SMALA web server via suitable views and can analyze the learning activities performed by individuals and groups of learners. 

Being a generic logging infrastructure, SMALA offers various interfaces for integrating concrete learning tools into the architecture. These interfaces are related to the authentication, definition, and handling of tool-specific semantic events. The following sections provide a detailed description of these interfaces and their usage.

\subsection{Software Components and their Interactions} 
The \textit{SMALA server} is a web server which, on the one hand, stores and displays log event streams and, on the other hand, is able to deliver the learning tools to the learners. In order to enable a learning tool to use the SMALA infrastructure, the tool must be registered with one or more \textit{learning activities} in the SMALA environment. This means SMALA knows about the tool and the corresponding activities via XML configuration files and is able to verify that it is allowed to use the SMALA logging service by an application specific key and a whitelist of allowed host URLs.
In principle, every learning tool based on web technologies, such as Java applets or AJAX web applications,
can be deployed and make use of SMALA's logging facilities.


Typically, SMALA-enabled learning tools are made available to the learners using an LMS (see Section~\ref{sec:log-views}) and thus, can be accessed as part of the learning materials available from the LMS course page. In order to use a learning tool, course participants only have to log into the LMS using their usual LMS user account and launch the tool. When starting up the tool, the LMS requests the creation of a new SMALA session for the current LMS user. This session is associated with all subsequent communication requests that occur between the learning tool and the SMALA logging service. As soon as a valid SMALA session is established, the learner can work with the tool in her local web browser. While doing so, all relevant interactions are wrapped as semantic events and sent to the logging service to store them in the database. Relevant interactions include both the user actions, the assessment results, and the feedback provided by the learning tools as a reaction to the users' activities. Identifying the relevant interactions is the responsibility of the learning tool designers and requires pedagogical expertise and a thorough understanding of the tool's application domain. Based on the SMALA log objects knowledge structure as described in the next section, the tool developers can model custom event objects as needed and add logging functionality to their tools by simply sending these events to the SMALA logging service.

The \textit{SMALA logging service} is implemented as a Java servlet and accepts all logging requests containing a valid SMALA session identifier and a serialized event object. The obtained event object is deserialized by the SMALA logging service and persisted to the SMALA logging database. Event objects have a {\textit type} and are associated with the current \textit{user session} and {\it learning activity}, i.e. a learning tool configured for a specific LMS course. All learning activities have to be registered with SMALA using XML configuration files and can have so-called {\it trigger classes} attached to them. If incoming events for these activities match the event type specified by these trigger classes, the triggers are activated and can handle the events appropriately. Currently, our semi-automatic assessment tools make use of this functionality for handling manual feedback events. When the SMALA logging service receives a manual feedback event object, it activates the configured {\it Send Mail trigger} which sends an email to the designated teacher or tutor, containing the learner's question, her email address, and the associated event data. So as to ensure the pseudonymity of the data, the email address is, then, removed from the event.

\begin{wrapfigure}{r}{80mm}\vspace{-22mm}
\includegraphics[width=8cm]{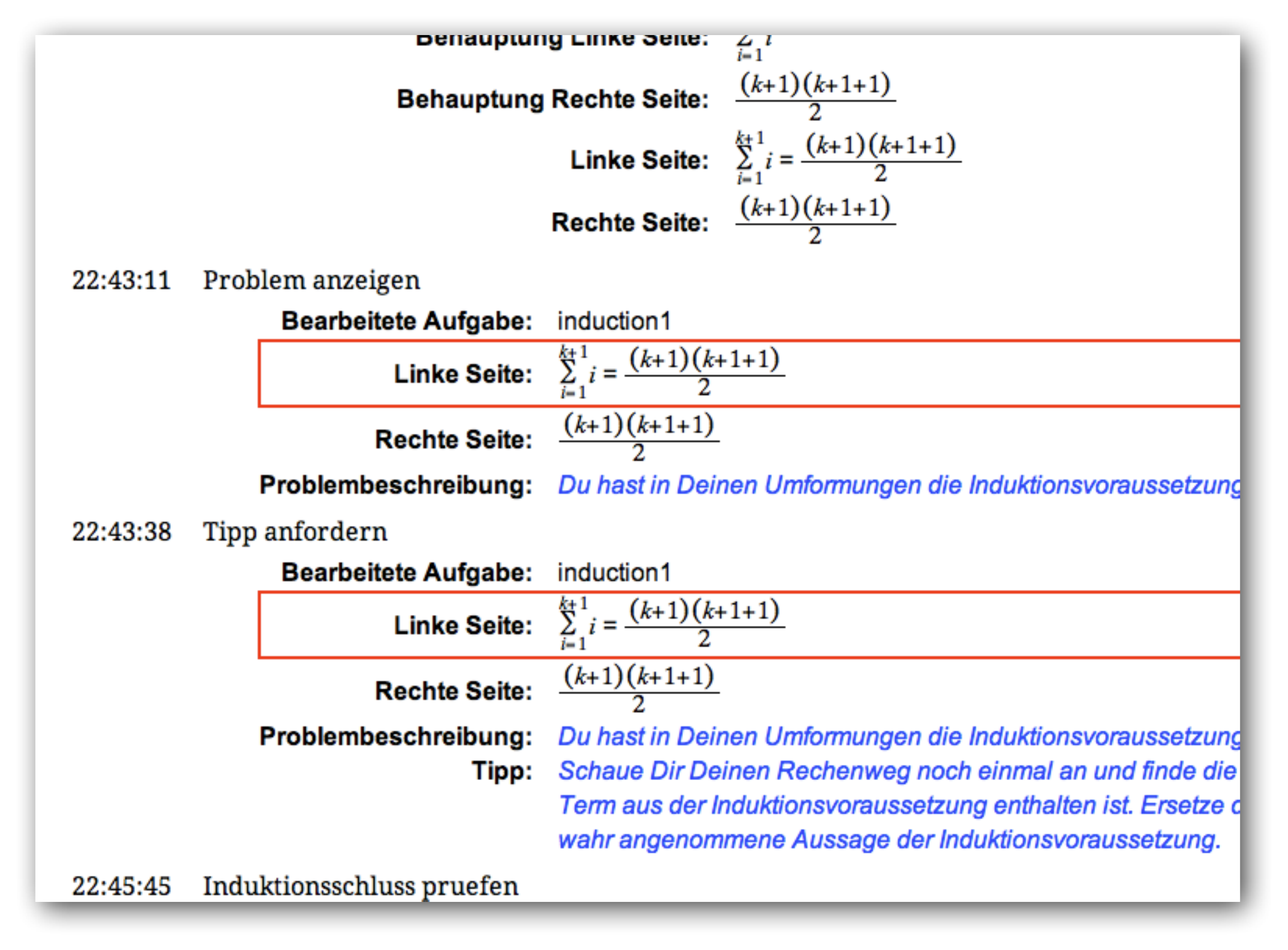}
\vspace{-10mm}\caption{The log view of a ComInM session.}
\label{fig:log-view-ComInM}
\vspace{-5mm}\end{wrapfigure}

Feeding events into the SMALA logging service is completely transparent to the end users of the learning tools. The learners' main focus is on using the learning tools and getting direct feedback from the tool's automatic assessment component. Teachers, however, can make use of SMALA's log views and optional tool-specific summary views on the recorded data (see section~\ref{sec:log-views}). These log views and summary views are presented using Java Server Pages (JSP) that query and render the event data from the SMALA logging data base. By allowing the deployment of tool-specific event renderer classes, not only \textit{general event data} such as the timestamp and a textual event description is displayed, but also \textit{tool-specific event data} such as rewritings, error messages or screenshots from the tool's user interface can be shown (see figure~\ref{fig:log-view-ComInM}). In the same way, tool-specific summary views and corresponding analyzer components for processing the semantic event data can be plugged into the SMALA infrastructure and integrated into the standard SMALA log views. 
The analysis done by these summary views offer the teacher a usable view to support his analysis and decision making.

SMALA offers a flexible \textit{authorization mechanism} to configure roles and access rules per activity: an LMS provenance and a signature 
enables an authorized {\it user source}; tutors (that can view log and deployment instructions) are authorized by obtaining identity from external identity providers such as Google and Facebook.

\subsection{Log Objects Knowledge Structure}
Log-event objects form the basic information entity that describes the users' actions.
The semantic-event-based data sets are grouped by {\it activity} and by {\it session} and form the starting point of the analytical process.

An activity is SMALA's concept for a learning tool that is offered from within a given course. It is configured with an authorization realm for teachers (a few external accounts) and for students (an integration method into an LMS). Each activity contains sessions which are a stream of log-events for a user.

Log-event objects have been designed with extensibility in mind since their semantics are very tool specific. 
The basic types include question events, image events, and action events. Common attributes of these basic types include the session-identifier, the timestamp, as well as the exercise name. Based on the exercise name the teacher can identify different parts of the learning activity. Tool makers can refine their types for each of the learning tools, extending the basic types described above.
The events can thus include the user's input (such as the OpenMath representations in the case of ComIn-M), the exercise state (such as the formula of the function being plotted in Squiggle-M), or even a URL to reconstruct the exercise state.


The serialization of the events transmitted to the server uses a generic XML format that is able to contain the tool specific logging information in form of key-value pairs, strings, numbers, dates and binary blobs, making the learning tools able to build their custom log event streams out of these datatypes. The events sent from the client are decoded on the server where they are validated and stored in the database using the Java Persistence Architecture.\footnote{The Java Persistence Architecture is an abstract API for object-relational mappings, see \url{http://jcp.org/aboutJava/communityprocess/final/jsr317/}.} This form of storage, close to the Java objects nature, allow sophisticated queries to be formulated so that log-views are carefully engineered to be relevant to the analysis of the students' activities for each learning tool.

\subsection{Availability}
SMALA is publicly available from \url{http://sail-m.de/sail-m/SMALA_en}.
Its source code (under the Apache Public License), technical documentation, and a link to the production and development servers with some demonstration parts are also linked there.

\section{Making the Learning Tools Available}\label{sec:teacher-deployment}

SMALA relies on a continuous identification of the users of the learning tools. The natural way to do so is to integrate the learnings tools in the LMS within the course page where students find their learning materials.

Each activity is established by the SMALA team: it is made by configurations of the roles, deployment explanations, and authorizations. A teacher then simply adds a link to the learning tool to his course page using code fragments suggested on the deployment explanations.
As a result, the learning tools can be started one or two clicks away from the main course page in the LMS, and thereby allow teachers to easily promote the tool during the lectures or on assignment sheets.


\section{Types of Log Views and their Usage}\label{sec:log-views}

Access to the logs starts with a {\it dashboard view} giving an overview of the recent activity with the learning tools. From there, one can get a list of users and a list of recent sessions. The users are not presented by name, but by their pseudonym, which we expect to be a number impossible to remember by teachers; 
each of them links to the list of sessions of that user, then to the detailed session-view. From the dashboard, one can also access overview pages giving a global analysis of the class performance.

\subsection{Session Views}
Sessions are displayed as a chronological sequence of user interactions in all detail. 
Each of the interactions' types are displayed with a custom representation. 

This enables the session log view to display, for example, 
each input the learner has made, and each fragment of feedback she has received. The purpose of this session view is to document the steps that were performed by the learner in the problem solving process and make it reproducible for the teacher.

This works well with learning tools that are made of dialogs where the learner submits an input and receives feedback.
An example is shown in the figure above: it shows a logging session using the learning tool ComIn-M where each of the induction proof constituents are presented. Such detailed log views enable lecturers to reproduce all actions a learner has made during the problem solving process.
However, this approach works less well for learning tools oriented towards direct manipulations such as Squiggle-M: log views then become long lists of small actions whose textual rendering is hard to read; e.g. {\it created point P1 in domain 1}, {\it created point P5 in domain 2}, ..., {\it linked P1 to P5}).

\newpage
\subsection{Requests for Help}
\begin{wrapfigure}{l}{70mm}
\vspace{-15mm}
\begin{center}
\includegraphics[width=70mm]{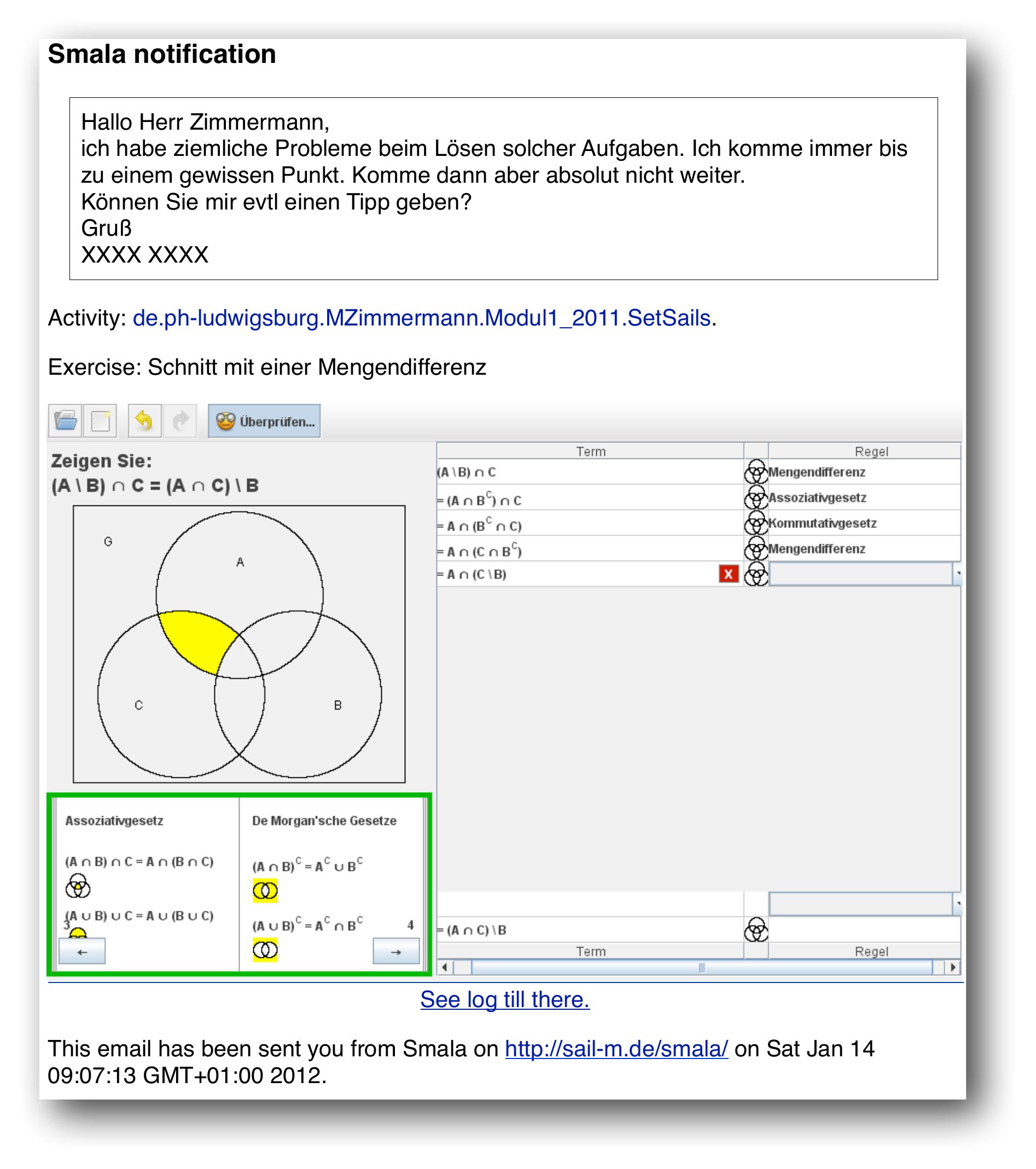}
\end{center}
\vspace{-8mm}
\caption{An actual tutor request email when using SetSails~\cite{Herding-Capture-and-replay} received during the evaluation.}
\label{fig:SetSails-Request}
\vspace{-7mm}
\end{wrapfigure}
A special type of interaction realizes the semi-automatic assessment approach~\cite{BHKMZ11}:
that of a request to the tutor formulated from within the learning tool. 
This interaction submits a special event to the SMALA server, containing the request of the student and a snapshot representing the current state of the tool (e.g. a screen copy). As soon as the SMALA server receives this event, the request is forwarded to the responsible teacher by email along with a URL to access the session until the moment of submitting the request.

These tutor request e-mails allow the teachers to grasp what the learner did (e.g. observing
that a function has been repeatedly used wrongly) therefore being able to  guide her accordingly. The semi-automatic feedback paradigm of~\cite{BHKMZ11} can be applied fully, enabling teachers to help students use the tool effectively and provide hints that go beyond the automated guidance of the learning tools.
An example tutor request email is shown in figure~\ref{fig:SetSails-Request}.

\begin{wrapfigure}{r}{67mm}\vspace{-15mm}\includegraphics[width=70mm]{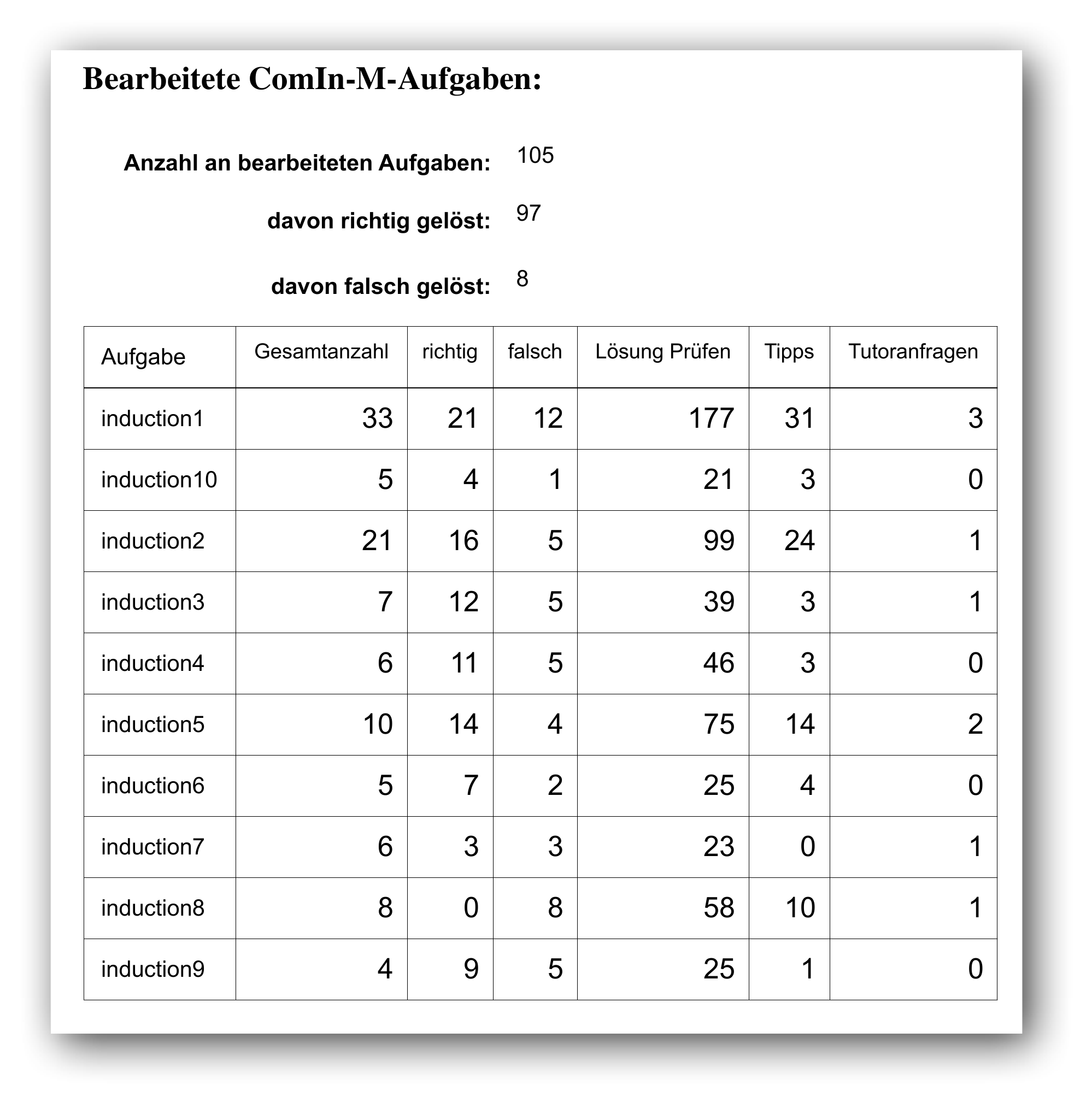}
\vspace{-10mm}\caption{A table of progress of the learners.}
\label{fig:ComInM-summary}
\vspace{-10mm}\end{wrapfigure}
\subsection{Summary Views}
The two log views of the previous sections support teachers to help individuals or get an idea of sample solution paths, but they are much less useful when trying to obtain a broader overview of a whole class's advances in the usage of the learning tools.

The most basic summary view is a list of all events of a given type. These allow teachers to find interesting sessions or detect a lack of achievement.
Another summary is provided by a dashboard with a timeline graph of the usage activity.

Another summary view is the display of the {\it success} or {\it failures} in the usages of the learning tools for each of the exercises that make up a learning tool. An example in figure~\ref{fig:ComInM-summary} shows a high involvement in attempting the first few exercises 
but clear difficulties for the remaining exercises as well as lack of attempts. The teacher could, from such a display, 
analyze and conclude a lack of understanding of the concepts, lack of engagement, or technical challenges; each of these hypothetical diagnoses could be answered by an in-class demonstration of success in one of the advanced exercises.

\section{Evaluation of Learning Tools using SMALA-Logging}
\vspace{-2mm}
This research has been made in parallel to several evaluations in which both the learning tools' qualities and the logging approach based on SMALA were evaluated.

An early evaluation was run in the summer of 2011 in order to get first feedback from the students and teachers at Karlsruhe University of Education: about forty students brought their laptops in two successive lectures and launched the MoveIt-M learning tool~\cite{Fest-Feedback} for exploring plane isometries and their composition. This first evaluation showed that obtaining a reasonable network bandwidth for launching the tool (20 MB download) may represent a challenge to other environments. Missing bandwidth may stimulate students to find workarounds such as choosing offline versions of the learning tools. Nonetheless, this first phase  showed that transmitting logs is technically easy.

Summative evaluations were run in the winter term of 2011-2012. In the basic mathematics courses of math teachers education three learning tools were deployed and used: Squiggle-M, SetSails~\cite{Herding-Capture-and-replay}, and ComIn-M. The logs witness the usage of the learning tools by $156$ users having run $965$ sessions yielding $24655$ events.
The logs have been complemented by a questionnaire filled out on paper and teacher interviews.


The evaluations have also shown that the transparent integration in which the learning tool is started by the learning management system and the logs are collected under pseudonym works fairly well and is acceptable to the students' privacy requirements. Indeed, only a small portion (less than $2~\%$) insisted on having no log collection.
Teachers that collaborated with us found it acceptable to expand the privacy circle when deploying such learning tools by enabling the learning management system's pages run by learners' browsers to transmit the user-information to the SMALA server. They made this decision because they acknowledged the usefulness and clarity of SMALA's mission.

\subsection{Feature Usage Statistics}

The central aspect of our evaluation was to measure the relevance of the semi-automatic assessment paradigm.
The usage of direct tutor requests from within the learning tools has been low (only 11 requests). 
Asked why they did not use the tutor request feature, many students responded that the automatic feedback had been sufficient (26\%). Most stated that they preferred asking peers (36\%) or tutors (26\%) directly (thus waiting till they had the chance to ask a question personnally), while a few responded that no help was necessary (11\%), or that formulating the question turned to be hard (15\%).
%
%

Although only few help requests were formulated (such as the one in figure~\ref{fig:SetSails-Request}), the requests that were actually submitted to the teachers could be answered: the detailed log views of the sessions have been sufficiently complete for them to answer questions with a precise hint that set the students on the right path. For this central workflow, the evaluation showed a successful setup.

A teacher who had seen the logs of ComIn-M commended the {\it screenshots of the users' input}. In fact, the logs do not contain screenshots, but MathML representations of the user's input. This comment shows that a faithful graphical display of the entire mathematical formulas is important to understand the students' solutions.

The usage of the logging feature for each of the teachers has been almost limited to responding to individual requests.
When interviewed at the end of the project, almost all teachers said they would consider using SMALA logging views more intensively if richer summary views were offered. 
Indeed, we focused on the log views to support individual requests, expecting overviews to be obtained by sampling a few sessions. As a result such summary views as the table in figure~\ref{fig:ComInM-summary} were implemented late in the development cycle. Our interviews also indicated that the separation of statistics {\it by exercise} seems to be an important basis of most log views.

Analyzing the web-server logs ($1813$ page views) showed that the students made use of the {\it myLog} links displayed aside of each tool launch. This feature is offered to support the transparency requirement. It allows each student to review her activities in the same format in which a teacher would see it.

\subsection{Technical Challenges}
The development of SMALA towards its sustained usage in day-to-day classrooms has been iterative, based on the feedback of the teachers and the students. It has faced the following technical challenges:

\paragraph{Genericity}
The SMALA server was designed to serve both web-applications and rich-client applications and it succeeded in doing so. Both ComIn-M (a web-server that sent all its logs to SMALA) and Java Applets (SquiggleM, MoveItM, and SetSails) could start and send identified logs to the SMALA server.
Although the learning tools send log-events of different types through different connection methods, the resulting log displays are delivered in a unified user interface.


\vspace{-3mm}\paragraph{Scalability} The choice of persistence layers such as the Java Persistence Architecture has been an important key to ensure the scalability to several thousands of log entries. For some display methods, large amounts of events still have to be inspected and it is still a matter of a few seconds (e.g. to display thousands of log entries). However, SMALA has sometimes been at the edge of performance even though it only handled a handful of courses.
For example, listing long sessions or listing all uploaded screenshots takes several minutes during which the data is extracted from the database and converted for display.
Summary views, in particular, need to strictly enforce the rule that their results are fetched using statistical queries to the database, which the tool developers enter using the JPQL language instead of iterating through numerous events.

An aspect at which the persistence libraries have shown to be too fragile is that of the evolution of the database schema of the log events. While creating an SQL schema based on the Java properties is transparent and effective (including support in IDEs to formulate queries), the persistence architecture offers almost no support to subsequently upgrade the SQL schema.

A modification such as raising the maximum size of a property (hence the size of an SQL column) needs either to be done manually in SQL or by an externalisation followed by a rebuild of the database. We ran the rebuild process at almost every deployment, the last taking about three hours. This process is, however, risky, and only thorough testing revealed that some information was lost in the process. The web nature of the log views, accessible by simple URLs allowed an easy test infrastructure: a simple web-page described each piece of information expected at each log-viewer URL. This allowed us to ensure that there are no unintended effects of the new versions' deployment.

\section{Conclusion}
In this paper we have described an approach to convey to practicing teachers what their students have been doing with learning tools. The approach relies on the usage of a software whose role is both to make available the learning tool in the learning management system and to display log views representing the learner's activity.
This approach turns the automatic feedback of the assessment tools into more intelligent tools which can exploit the teacher's knowledge.
The implemented toolset conveys broad enough information to watch the learners' overall progress, and to respond to individual queries.

Below we present research questions that this approach opens.

\subsection{Open Questions}

The involvement of the teachers in the usage of the learning tools remains a broad area of research which this paper contributes to. It is a subtle mix of mathematical competency, technical skills, and pedagogical sense -- all of which are addressed by a teacher support tool. It is bound to the teachers' representations of the mathematical knowledge that students have in mind, the meaning they attach behind them and behind the operations with them.

The {\it instrumental orchestration}~\cite{Drijvers-Trouche-Orchestra} is a model of the role of the teacher who uses the instruments (the tools) to let himself and the students (the orchestra) play the mathematical music. It mostly studies the usage of the learning tools in class, while students often use learning tools in the comfort of their homes: the learning tools are there, precisely, to let the learners deepen or discover at their own pace, away from the tight rythm of the lectures.

The research of this paper contributes to letting the teacher perceive this activity.
In this paper we have proposed a few solutions relying on both detailed and summary views. Does it imply a high level of diagnostics?
The survey~\cite{Zinnetal-StudentDistance-ectel-2006} indicates that higher level indicators such as the learners' mastery level and misconceptions are much desired. We note, however, that they would only be useful if they can be trusted by teachers to be complete and precise; we are not sure that this is easily achievable, and indeed the paper highlights this difficulty in its conclusion. Sticking to more concrete events, such as {\it typical feedback types} might be easier to implement, easier to represent to the teacher, directly linkable to steps in session actions, and thus much more trustable.


We have introduced SMALA in relatively traditional didactical settings in which the learning tools are mostly used for the purposes of exercising. This is an important aspect but other didactical setups can leverage a logging architecture.

Among didactical setups is the exploitation of the {\it own log} feature beyond the mere transparency requirement. Displaying the individual logs of a selected learner, this feature can be used by the student as a personal log to be reminded of the competencies she has already acquired, much similar to the learning-log approach. What needs to be achieved to encourage and make possible such practice? Should links be made from a personal diary into concrete sessions?


As another example of a different didactical setup, Erica Melis suggested to display the logs of past uses of the learning tool overlaid on the {\it behaviour graph} of the Cognitive Tutor Authoring Tools~\cite{CTAT-NewParadigm-AuthoringITS} and to use this in coordination with the learning tools. This approach would constitute a discussion tool that allows the teacher to explain the possible avenues to solve the exercise, changing the state of the exercise and discussing what was correct or what was wrong. An approach and architecture such as SMALA makes such a tool possible.

\paragraph{Acknowledgements}
This research has been partly funded by the Ministry of Education and Research of Germany in the SAiL-M project.
We thank our student assistants Torsten Kammer and Michael Binder for their active support.

%
%
\newcommand{\etalchar}[1]{$^{#1}$}

\end{document}